\documentclass[conference]{IEEEtran}
\usepackage{tabularx}
\usepackage{ifpdf}
\usepackage{graphicx}
\usepackage{glossaries}
\usepackage{amssymb}
\usepackage{amsmath}
\usepackage{lipsum}
\usepackage{cite} 
\usepackage{epstopdf}
\usepackage{mathrsfs}
\usepackage{cite}
\usepackage{color}
\usepackage{cuted}

\usepackage[compact]{titlesec}       

\usepackage[justification=centering]{caption}
\usepackage[keeplastbox]{flushend}

\newcommand{\lb} {\left}
\newcommand{\rb} {\right}
\newcommand{\nn} {\nonumber}
\makeatletter
\def\ScaleIfNeeded{
\ifdim\Gin@nat@width>\linewidth \linewidth \else \Gin@nat@width \fi
} \makeatother

\newcommand{\xd}{\nonumber \\}

\newcommand{\Ei}{\mathrm{Ei}}

\begin{document}
 \onecolumn{\noindent © 2021 IEEE. Personal use of this material is permitted. Permission from IEEE must be obtained for all other uses, in any current or future media, including reprinting/republishing this material for advertising or promotional purposes, creating new collective works, for resale or redistribution to servers or lists, or reuse of any copyrighted component of this work in other works.}
 \twocolumn{
\title{Transmitter Selection for Secrecy in Cognitive Small-Cell Networks with  Backhaul  Knowledge}
\author{
    \IEEEauthorblockN{Burhan Wafai\IEEEauthorrefmark{1}, Chinmoy~Kundu\IEEEauthorrefmark{2}, 
    Ankit Dubey\IEEEauthorrefmark{3},
    Jinghua Zhang\IEEEauthorrefmark{4}, 
    and Mark F. Flanagan\IEEEauthorrefmark{5},
    }
    \IEEEauthorblockA{\IEEEauthorrefmark{1}\IEEEauthorrefmark{3}Department of EE, Indian Institute of Technology Jammu, Jammu \& Kashmir, India }
    
     \IEEEauthorblockA{\IEEEauthorrefmark{4}School of Electronics, Electrical Engineering and Computer Science, Queen’s University Belfast, U.K.}
     \IEEEauthorblockA{\IEEEauthorrefmark{2}\IEEEauthorrefmark{5}School of Electrical and Electronic Engineering, University College Dublin, Belfield, Ireland}
    \textrm{\IEEEauthorrefmark{1}burhanwafai@ieee.org}, {\IEEEauthorrefmark{2}chinmoy.kundu@ucd.ie},{\IEEEauthorrefmark{3}ankit.dubey@iitjammu.ac.in}, {\IEEEauthorrefmark{4}jzhang22@qub.ac.uk},
     {\IEEEauthorrefmark{5}mark.flanagan@ieee.org} }

\maketitle
 \thispagestyle{empty}
 \pagestyle{empty}
\pagestyle{plain} 
\begin{abstract}
A small-cell network with multiple transmitters and unreliable wireless backhaul is considered for secrecy enhancement. The small-cell network is operating under a spectrum sharing agreement with a primary network in a cognitive radio system. A constraint on the desired outage  probability at the primary receiver is assumed as a part of the spectrum sharing agreement. The reliability of the wireless backhaul links are modeled by a set of independent and identically distributed Bernoulli random variables. A sub-optimal and an optimal small-cell transmitter selection (TS) scheme is proposed to improve the performance of the system, depending on the availability of channel state information. Selection schemes are designed for the scenario where knowledge is available regarding which backhaul links are active. The corresponding secrecy outage probabilities along with their asymptotic expressions are derived. It is shown that the secrecy performance is significantly improved compared to the case where knowledge of the active backhaul links is unavailable. 
\end{abstract}
\vspace{-.5cm}
\section{Introduction}\label{section 1}
The emerging Internet of Things (IoT), sometimes also referred to as the Internet of Everything, will be deployed in future wireless networks for massive connectivity \cite{Jeffrey20145G}. In the fifth-generation (5G) and beyond wireless technologies, the networks will be dense and consist of heterogeneous small cells {\cite{wong2017key,chia2009next}}, which will share the same spectrum in a cognitive radio (CR) fashion \cite{Mitola1999Cognitive} to meet the demands of high data rate applications in the future IoT. Furthermore, the dense network deployment will lead to dense backhaul connections from the backbone to multiple small-cell transmitters. Although the traditional wired backhaul offers high reliability and data rate, the deployment cost is unsustainable in 5G and beyond large-scale networks {\cite{Orawan2011Evolution,Xiaohu20145G, coldrey2012small}}. As an alternative to wired backhaul, wireless backhaul has emerged as a cost-effective and flexible solution. In the future, the wireless backhaul deployment can form the backbone link between a macro-cell and many small cells in a CR based heterogeneous network (HetNet). The intrinsic shortcomings of wireless backhaul, i.e., non-line-of-sight (NLOS) propagation and multipath fading, have prompted investigations towards improving the quality of transmission with wireless backhaul \cite{kim2017performance,khan2015performance}. 

CR is an effective way to alleviate inefficient frequency usage and spectrum scarcity. To ensure simultaneous transmission by unlicensed secondary users and licensed primary users, a possible way is to use a CR underlay scheme. As such, frequency sharing using a CR network is essential to increase the capacity and spectral efficiency of the system.
The authors in \cite{huy2017cognitive} studied the impact of unreliable backhaul links on cooperative HetNets in a CR environment. The authors in \cite{nguyen2017multiuser} proposed various relay selection schemes and multiuser scheduling for cognitive networks with unreliable backhaul links. The authors in \cite{nguyen2017cognitive} studied the performance of the best relay selection scheme in cognitive HetNets in the presence of unreliable backhaul connections. However, the authors in \cite{nguyen2017multiuser,huy2017cognitive, nguyen2017cognitive} neglected the  interference to the secondary network and  only considered the interference to the primary network, without any secrecy constraint.

Due to the broadcast nature of the wireless channel, any  unintended  receiver can potentially access the confidential information in the transmission. Due to the dynamic and ubiquitous nature of the network architecture, HetNets with small-cell architecture are particularly prone to eavesdropping. The secrecy performance of cooperative single-carrier HetNets  with unreliable wireless backhaul connections was investigated in  \cite{nguyen2017secure}.  The secrecy performance of an energy harvesting relay network with unreliable wireless backhaul was studied in \cite{Yincheng}, where secrecy is enhanced by different transmitter selection (TS) schemes. The effect of  wireless backhaul connections on secrecy performance of finite-sized cooperative systems for multiple passive eavesdroppers was explored in \cite{kim2016secrecy}. Even though the consequences of unreliable wireless backhaul on secrecy has been considered in the aforementioned literature,  little work along this line exists for CR networks. 

Recently, \cite{Vu2017Secure}, and \cite{Kundu_TVT19} investigated the effect of wireless backhaul on secure communication techniques in CR networks. Wireless backhaul reliability for security in a CR network was studied in \cite{Vu2017Secure} but the system model was simplified by neglecting interference at the eavesdropper and destination. The authors in \cite{Kundu_TVT19} studied secrecy improvement of a CR network with wireless backhaul while guaranteeing the Quality-of-Service (QoS) of the primary network. However, these secrecy protocols did not consider whether backhaul was active or not. Assuming that the backhaul activity knowledge is available to the TS procedure, a recurrent neural network was used for optimal TS in \cite{ShaliniGC20} to improve the secrecy outage probability (SOP) of the system of \cite{Kundu_TVT19}, however, neither a closed-form solution nor an asymptotic analysis was provided. Here we improve the secrecy performance of \cite{Kundu_TVT19} by designing TS protocols that actively take into account the knowledge of active backhaul links and provide closed-form solutions as well as asymptotic analysis.

 In this paper, we  consider  a CR  network in a spectrum sharing environment with  an access point (AP) providing a wireless backhaul connection to multiple secondary small-cell transmitters. 
 To mitigate the effect of eavesdropping on the secondary transmission, we propose selection diversity to improve secrecy by choosing the best transmitter. However, in contrast to \cite{Kundu_TVT19}, it is assumed that the knowledge of the activity status of the backhaul links is available before the transmitter is selected. This indicates that the best transmitter can be selected among the transmitters with active backhaul links. This will further improve the performance over \cite{Kundu_TVT19} which does not consider activity knowledge for TS.  Our main contributions are listed as follows: 
\begin{itemize}
\item Depending on the level of available channel state information (CSI) at the transmitters, we  propose two TS schemes: (i) a sub-optimal scheme for transmitter selection (STS), in which the channel gain from  the  secondary  transmitter  to  the  destination is maximized; and (ii) an optimal scheme for transmitter selection (OTS), in which the secondary secrecy capacity is maximized.  
\item The secrecy outage probability (SOP) for STS is obtained in closed-form, while a computable expression is derived for the OTS scheme.
\item To obtain greater insights, an asymptotic  analysis is also included, which provides a closed-form solution (valid in the high-SNR regime) for the SOP for each of the proposed schemes.  
\end{itemize}  

The rest of this paper is organised as follows. The  system  and  channel  models are described in Section~\ref{section 2}. The small-cell TS schemes are proposed and the corresponding secrecy outage probabilities (SOPs) are evaluated in Section~\ref{section 3}, while the asymptotic secrecy expressions are derived in Section~\ref{section 4}. The numerical results  are presented with discussion in Section~\ref{section 5}, and finally conclusions are drawn in Section~\ref{sec_Conclusion}.

\textit{Notation:} The probability of occurrence of an event is represented by $\mathbb{P}[\cdot]$, $\mathbb{E}_X[\cdot]$ denotes expectation of a random variable (RV) $X$. $F_{X} (\cdot)$ represents the cumulative distribution function (CDF) of $X$, and $f_{X} (\cdot)$ is the corresponding probability density function (PDF). For any two nodes $A$ and $B$, the channel coefficient is denoted by $h_{AB}$ and the signal-to-interference-plus-noise-ratio (SINR) at $B$ is denoted as $\Gamma_{AB}$. 
\section{System and channel model}\label{section 2}
\begin{figure}
 \centering 
 \includegraphics[width=2.5in]{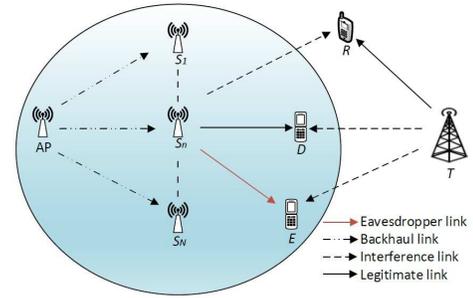} 
  \vspace{-.21cm}
 \caption{
 Cognitive radio network with wireless backhaul.}
 \label{fig:SM}
 \vspace{-.5cm}
 \end{figure}
The system consists of a primary network and a cognitive secondary network which shares the spectrum of the primary network as shown in Fig. \ref{fig:SM}. The primary network consists of a transmitter,  $T$, and a receiver, $R$, whereas the secondary network includes $N$ small-cell transmitters, $S_n$, $ n=1, \ldots, N$  serving a user $D$. The small-cell transmitters are connected to an AP via wireless backhaul links.  An eavesdropper $E$ is present in the system to intercept the secondary transmission. Due to the concurrent primary and secondary transmissions, interference  from the primary affects the secondary reception and vice versa. It is assumed that the channel coefficient between each pair of nodes follows an independent Rayleigh distribution, and thus the corresponding power gains are independent and exponentially distributed. For each $n=1, \ldots, N$, the links $S_n$-$D$,  $S_n$-$E$ and $S_n$-$R$ have independently distributed power gains with exponential distribution parameters $\lambda_{sd}$, $\lambda_{se}$, and $\lambda_{sr}$.
The noise at each receiver is modeled as complex additive white Gaussian noise (AWGN) with  zero mean and  variance $N_0$. 

Each wireless backhaul link from AP to small-cell transmitter has a certain probability of failure. The reliability of the backhaul links is modeled by independent identically distributed Bernoulli RVs, where the probability of each link being active is $s$ and the probability of a link being inactive is $1-s$.
With the knowledge of backhaul activity, TS is carried out before data transmission begins.  It is assumed that all relevant instantaneous CSI, including which backhaul links are active, is available at the transmitters for making the appropriate TS. The system model and proposed TS schemes are similar to those presented in \cite{Kundu_TVT19}; however, the main difference is that here TS is carried out with the knowledge of active backhaul links, whereas no knowledge of backhaul activity was assumed before selection in \cite{Kundu_TVT19}. For this reason, the derivation of the SOP and its asymptotic expression are not straightforward based on the results in \cite{Kundu_TVT19}.

\subsection{Interference from the Primary Transmitter}
The concurrent transmission induces interference in secondary receivers due to the primary transmission. 
Hence, conditioned on the selected transmitter, the SINR at $E$ can be expressed as
\begin{align}
\label{SNR_E}
\Gamma_{S_nE}&=\dfrac{P_S  |h_{S_{n}E}|^2}{P_T  |h_{TE}|^2+N_0}\,,
\end{align}
where  $P_T$ is the transmitted power at $T$  and $P_S$ is   allowable 
transmit power at $S_n$. The expression of $\Gamma_{S_nD}$ is similar to that of  $\Gamma_{S_nE}$, but with subscripts $E$ replaced by $D$.
\subsection{Secondary Transmit Power Constraint}
A power constraint is set on the secondary transmitter to guarantee a certain primary QoS during simultaneous secondary transmission. We consider that the primary outage probability should always be below a threshold outage probability $\Phi$, i.e.,
\begin{align}\label{desired outage probability}
\mathbb{P}\left[{\Gamma_{TR}}  < \Gamma_0\right]\leq \Phi,
\end{align}
where $0 < \Phi <1$ and $\Gamma_0 = 2^{\beta}-1$ where $\beta$ represents the threshold rate for the primary network. We will proceed to find the CDF of $\Gamma_{TR}$ in order to find the allowable $P_S$. 

As the primary receiver is interfered by the selected transmitter, $\Gamma_{TR}$ can be expressed similar to (\ref{SNR_E}) as 
\begin{align}\label{SINR at P_R}
\Gamma_{TR}= \frac{P_T|h_{TR}|^2}{P_S|h_{S_nR}|^2+N_0}.
\end{align}
Its CDF is then evaluated as 
\begin{align} 
\label{CDF_TR}
&F_{\Gamma_{TR}}(x)=\mathbb{P}\lb[\Gamma_{TR}\le x\rb]\nn \\
&=\mathbb{P}\left[|h_{TR}|^2\le\frac{x\left(P_S|h_{S_nR}|^2+N_0\right)}{P_T }\right] \nn \\
&=\int_{0}^{\infty}\left[1-\exp\left(\frac{-\lambda_{tr}(P_Sy+N_0)}{P_T}x\right)\right] 
\lambda_{sr}\exp\left(-\lambda_{sr}y\right)dy \nn \\
&=1-\frac{\frac{\lambda_{sr}\Gamma_T}{\lambda_{tr}\Gamma_{S}}}{x+\frac{\lambda_{sr}\Gamma_T}{\lambda_{tr}\Gamma_{S}}}\exp\left(\frac{-\lambda_{tr}x}{\Gamma_T}\right),
\end{align} 
where $\Gamma_T=P_T/N_0$ and $\Gamma_S=P_S/N_0$.
The constraint on $P_S$ can be evaluated by combining (\ref{desired outage probability}) and (\ref{CDF_TR}) as
\begin{align} \label{xi_upper}
&F_{\Gamma_{TR}}(\Gamma_0)\le\Phi.
\end{align}
After some simple manipulations, this leads to the constraint 
\begin{align}\label{P_S}
    P_S=\left\{
                \begin{array}{ll}
                  P_T\lambda_{sr}\xi\ \ \ \textnormal{if}\ \ \ \xi>0\\
                   0\ \ \ \ \ \ \ \ \ \ \ \ \textnormal{otherwise},\\
                \end{array}
              \right.
\end{align}
where
\begin{align} \label{xi}
\xi=\frac{1}{\lambda_{tr}\Gamma_0}\left[\frac{\textnormal{exp}\left(\frac{-\lambda_{tr}\Gamma_0}{\Gamma_{T}}\right)}{1-\Phi}-1\right].
\end{align}
We find from (\ref{P_S}) and (\ref{xi}) that $P_S$ is proportional to the secondary to primary channel quality and inversely proportional to the primary channel quality, respectively.

\section{SOP of Secondary TS Schemes} 
\label{section 3}
In this section we will propose a sub-optimal and an optimal secondary TS scheme, considering that backhaul activity knowledge is  \textit{a priori} available, and determine the corresponding SOPs.
The SOP of the secondary network is defined as the probability that the secrecy rate, $C_{S}$, of the secondary network falls below the threshold secrecy rate, $R_{th}$, and is given as 
\begin{align}\label{secrecy OP equation}
\mathcal{P}_{out}&= \mathbb P\left[C_{S} < R_{th}\right]=\mathbb{P}\left[\frac{1+\Gamma_{SD}}{1+\Gamma_{SE}}<\rho\right]
\nonumber \\
&= \int_{0}^{\infty} F_{\Gamma_{SD}}(\rho(x+1)-1) f_{\Gamma_{SE}}(x) dx,
\end{align}
where 
$\Gamma_{SD}$ and  $\Gamma_{SE}$ are the equivalent SINRs of the destination and eavesdropper links, respectively, when the TS scheme is applied, $\rho = 2^{R_{th}}$.
Note that the expression for the SOP in  (\ref{secrecy OP equation}) uses the fact that $\Gamma_{SD}$ and $\Gamma_{SE}$ are independent. We will first derive  the distributions $F_{\Gamma_{SD}}(\cdot)$ and $f_{\Gamma_{SE}}(\cdot)$ for each selection scheme, and then apply the resulting expressions in order to evaluate (\ref{secrecy OP equation}).   The selection schemes and their corresponding SOPs are derived in the following subsections. 

\subsection{Sub-optimal Scheme for  Transmitter Selection (STS)}
In the STS scheme, the transmitter is selected (among those with active backhaul links) which has maximum power gain to the destination. 
The equivalent channel power  gain becomes
\begin{align} 
\label{eq_gain_max}
\gamma_{SD}=\max_{n \in \mathcal{S}}|h_{S_n D}|^2 = \max_{{n \in \{1,\ldots, N\}}}|\hat{h}_{S_n D}|^2,
\end{align}
where $\mathcal{S}\subseteq\{1,2,\ldots,N\}$ denotes the subset of transmitters for which the backhaul link is active and $|\hat{h}_{S_n D}|^2$ models the power gain of the $S_n$-$D$ link together with its corresponding backhaul link. $\hat{h}_{S_n D} = h_{S_n D}$ if $n \in \mathcal{S}$, and $\hat{h}_{S_n D} = 0$ otherwise.
Hence, the destination SINR becomes
\begin{align}\label{SINR_SD}
\Gamma_{SD}=\frac{ P_S \gamma_{SD}}{P_T|h_{TD}|^2+N_0}.
\end{align}
We will first find out the distribution of $\gamma_{SD}$ and subsequently the distribution of $\Gamma_{SD}$ and $\Gamma_{SE}$ for (\ref{secrecy OP equation}). 

As backhaul uncertainty is modelled using a Bernoulli RV, the PDF of the channel power gain including backhaul link, $|\hat{h}_{S_nD}|^2$,  for each $n\in\{1,\ldots,N\}$, has the mixture distribution
\begin{align} \label{mixture_dist}
f_{|\hat{h}_{S_nD}|^2}(x)=(1-s)\delta(x) + s f_{|h_{S_nD}|^2}(x),   \end{align}
where $\delta(x)$ is the delta function, $|h_{S_nD}|^2$ is the power gain of link $S_n$-$D$ and $s$ is the probability of each link being active.
If no backhaul links are active, it is considered that a secrecy outage has occurred. 
The CDF of (\ref{mixture_dist}) can be written  as
\begin{align} \label{eq_cdf_mixture_dist}
F_{|\hat{h}_{S_nD}|^2}(x)=1-s+sF_{|h_{S_nD}|^2}(x)=1-s\exp\lb(-\lambda_{sd}x\rb).    
\end{align}
Now the distribution of $\gamma_{SD}$ in (\ref{eq_gain_max}) including backhaul uncertainty can be evaluated  with the help of (\ref{eq_cdf_mixture_dist}) as
\begin{align}
& F_{\gamma_{SD}}(x)=\mathbb{P}\lb[\gamma_{SD}<x\rb]=\mathbb{P}\lb[\max_{n \in \{1,\ldots, N\}} |\hat{h}_{S_nD}|^2 \le x\rb]\nn\\
&=\lb(1-s\exp\lb(-\lambda_{sd}x\rb)\rb)^N 
\nn \\
&=1-\sum_{n=1}^{N}\binom{N}{n}(-1)^{n+1}s^n\exp\left(-n\lambda_{sd}x\right) .
\end{align}
Note that in contrast to \cite{Kundu_TVT19}, the modeling of $F_{\gamma_{SD}}$ will ensure that the selected transmitter is the one whose backhaul link is active. 
Next we concentrate on finding the CDF of ${\Gamma}_{SD}$  and PDF of ${\Gamma}_{SE}$. The derivation of the CDF of  ${\Gamma}_{SD}$ in (\ref{SINR_SD}) including backhaul reliability can be obtained with the help of $F_{\gamma_{SD}}(x)$ as
\begin{align}
\label{CDF_SNR_SD}
&F_{\Gamma_{SD}}(x)=\mathbb{P}\left[\Gamma_{SD}\le x\right]
=\mathbb{P}\left[ 
\gamma_{SD}\le\frac{(\Gamma_T|h_{TD}|^2+1)x}{\Gamma_S}\right]\nn\\ 
&=\int_{0}^{\infty}
\left[1-
\sum_{n=1}^{N}\binom{N}{n}(-1)^{n+1}s^n \rb.\nn \\
&\lb.\times\exp\left(\frac{-n\lambda_{sd}(\Gamma_Ty+1)x}{\Gamma_S}\right)\right]\lambda_{td}\exp\left(-\lambda_{td}y\right)dy
\xd
&=1-\sum_{n=1}^{N}\binom{N}{n}\frac{(-1)^{n+1} 
\frac{\lambda_{td}\Gamma_{S}}{n\lambda_{sd}\Gamma_{T}}}{x+
\frac{\lambda_{td}\Gamma_{S}}{n\lambda_{sd}\Gamma_{T}}}s^n\exp\left( \frac{-n\lambda_{sd}x}{\Gamma_{S}}\right).
\end{align}


Note that the transmitter selection is independent of the channel from the secondary transmitter to eavesdropper. Thus, the CDF of $\Gamma_{SE}$ conditioned on the selected transmitter can be directly obtained by following (\ref{CDF_TR}) as
  $\Gamma_{SE}=\Gamma_{S_nE}$, i.e., 
\begin{align}\label{CDF_SNR_E}
&F_{\Gamma_{SE}}(x)=
1-\frac{\frac{\lambda_{te}\Gamma_{S}}{\lambda_{se}\Gamma_{T}}}{x+\frac{\lambda_{te}\Gamma_{S}}{\lambda_{se}\Gamma_{T}}} \exp\left(-\frac{\lambda_{se}x}{\Gamma_{S}}\right),
\end{align}
and by differentiating the above CDF the corresponding PDF can be obtained as \begin{align}\label{PDF_SNR_E}
&f_{\Gamma_{SE}}(x)=\frac{\frac{\lambda_{te}}{\Gamma_{T}}\exp\left(-\frac{\lambda_{se}x}{\Gamma_{S}}\right)}{x+\frac{\lambda_{te}\Gamma_{S}}{\lambda_{se}\Gamma_{T}}}+\frac{\frac{\lambda_{te}\Gamma_{S}}{\lambda_{se}\Gamma_{T}}\exp\left(-\frac{\lambda_{se}x}{\Gamma_{S}}\right)}{\left(x+\frac{\lambda_{te}\Gamma_{S}}{\lambda_{se}\Gamma_{T}}\right)^2}.
\end{align}

Next we substitute $F_{\Gamma_{SD}(\cdot)}$ and $f_{\Gamma_{SE}(\cdot)}$ from (\ref{CDF_SNR_SD}) and (\ref{PDF_SNR_E}), respectively,  into (\ref{secrecy OP equation}) to obtain the SOP in closed-form as
\begin{align}\label{SOP_STS}
\mathcal{P}_{out}&=
1-\sum_{n=1}^{N}\binom{N}{n}(-1)^{n+1}\frac{\lambda_{te}\lambda_{td}\Gamma_{S}}{n\rho\lambda_{sd}\Gamma_{T}^2}s^n\nn \\ 
&\exp\left(-\frac{n\lambda_{sd}\left(\rho-1\right)}{\Gamma_{S}}\right)\lb(I_1+\frac{\Gamma_{S}}{\lambda_{se}}I_2\rb),
\end{align}
where 
\begin{align}
\label{I_1}
I_1=&\frac{\exp\left(ac\right)\Ei\left(-ac\right)}{a-b}
-\frac{\exp\left(bc\right)\Ei\left(-bc\right)}{a-b},
\end{align} and
\begin{align}
\label{I_2}
I_2=&-\frac{\exp\left(ac\right)\Ei\left(-ac\right)}{\left(a-b\right)^2}
+\frac{\exp\left(bc\right)\Ei\left(-bc\right)}{\left(a-b\right)^2}
\xd
+&\frac{\left(c\exp\left(bc\right)\Ei\left(-bc\right)+\frac{1}{b}\right)}{a-b}, 
\end{align}
with 
$a=\frac{\lambda_{td}\Gamma_{S}+n\rho\lambda_{sd}\Gamma_{T}-n\lambda_{sd}\Gamma_{T}}
{n\rho\lambda_{sd}\Gamma_{T}}$, $b=\frac{\lambda_{te}\Gamma_S}{\lambda_{se}\Gamma_T}$, and
$c=\frac{n\rho\lambda_{sd}+\lambda_{se}}{\Gamma_S}$. 

 
\subsection{Optimal Scheme for Transmitter Selection (OTS)}
The STS scheme described in the previous subsection uses only the destination channel knowledge, which is sub-optimal. If global channel state information is available, we can determine the transmitter for which the instantaneous achievable secrecy rate of the secondary network is maximum, via  
\begin{align}\label{OTS_CS}
n^* = \max_{n\in \mathcal S} \{ {C^{n}_S}\},
\end{align}
where for each $n$, $C^{n}_S$ refers to the wiretap channel formed by the $S_n$-$D$ and $S_n$-$E$ links. 

For each $n$, the secrecy capacity $C_S^n$ depends on the common links $T$-$D$ and $T$-$E$ of the individual wiretap channel; hence, the derivation of the optimal SOP considers the conditional SOP with respect to these two RVs first and then finally averages over these. Including backhaul activity knowledge for selection, the SOP can be evaluated as  
\begin{align}
\label{sop_OTS}
\mathcal{P}_{out}&=\mathbb P\left[ \max_{n\in\{1, \ldots, N\}}\{C_{S}^n\} < R_{th}\right] \nn \\
&=\mathbb E_{|h_{TD}|^2 }\mathbb E_{|h_{TE}|^2}\left[\mathbb P \lb[C^n_{S}\le R_{th}\bigg{|}|h_{TD}|^2, |h_{TE}|^2\rb]\right]^N \nn \\
&=\int_0^\infty\int_0^\infty 
\lb[\int_0^\infty F_{\Gamma_{SD}}(\rho(t+1)-1|x) f_{ \Gamma_{SE}}(t|y)dt\rb]^N \nn \\
&\times \lambda_{td}\exp\left(-\lambda_{td}x\right)\lambda_{te}\exp\left(-\lambda_{te}y\right)dxdy\nn\\
&=\int_0^\infty\int_0^\infty 
\lb[1-\frac{\lambda_{se}\left(\Gamma_Ty+1\right)s}{\rho\lambda_{sd}\left(\Gamma_Tx+1\right)+\lambda_{se}\left(\Gamma_Ty+1\right)}\rb.\nn \\
&\lb.\times\exp\left(\frac{-\lambda_{sd}\left(\rho-1\right)\left(\Gamma_T x+1\right)}{\Gamma_S}\right)
\rb]^N \nn \\
&\times \lambda_{td}\exp\left(-\lambda_{td}x\right)\lambda_{te}\exp\left(-\lambda_{te}y\right)dxdy.
\end{align}
where $F_{\Gamma_{SD}}(\cdot|x)$ and $f_{\Gamma_{SE}}(\cdot|y)$ are the
conditional CDF and PDF of $\Gamma_{SD}()$ and  $\Gamma_{SE}()$ conditioned on $|h_{TE}|^2$ and $|h_{TD}|^2$, respectively. $F_{\Gamma_{SD}}(\cdot|x)$ and $f_{\Gamma_{SE}}(\cdot|y)$ are derived following a similar method of (\ref{CDF_SNR_SD}) and (\ref{PDF_SNR_E}), respectively, for the conditional case. $F_{\Gamma_{SD}}(\cdot|x)$ includes the backhaul parameter $s$.
The above double integral does not admit a closed-form solution and is therefore evaluated using Mathematica software.
\section{Asymptotic Secrecy Analysis}\label{section 4}
In this section, the asymptotic analysis (i.e., assuming $\Gamma_T  \rightarrow \infty$) of considered TS schemes will be performed, in order to  provide a better insight into the impact of unreliable backhaul connections. 

\subsection{Sub-optimal Scheme for Transmitter Selection (STS)}
As  $\Gamma_T \rightarrow \infty$, $\Gamma_S$ also tends to $\infty$, and thus
$\textnormal{exp}\lb(\frac{-n\lambda_{sd}(\rho-1)}{\Gamma_S}\rb)$ tends to unity; it follows that $\sum_{n=1}^{N}\binom{N}{n}(-1)^{n+1}\frac{\lambda_{te}\lambda_{td}\Gamma_{S}}{n\rho\lambda_{sd}\Gamma_{T}^2}s^n I_1$ tends to zero. By substituting the value of $\Gamma_S$ from (\ref{P_S}) and assuming $\Gamma_T  \rightarrow \infty$, the asymptotic expression for the SOP in 
(\ref{SOP_STS}) can be approximated as 
\begin{align}\label{SOP_STS_asym}
\mathcal{P}_{out}&\approx1-\sum_{n=1}^{N}\binom{N}{n}(-1)^{n+1}\frac{\lambda_{te}\lambda_{td}\lambda_{sr}^2\xi^2}{n\rho\lambda_{sd}\lambda_{se}}s^n I_2 
\end{align}
where $\xi$ from (\ref{xi}) is approximated for $\Gamma_T  \rightarrow \infty$  as
\begin{align} \label{xi_STS_asym}
    \xi\approx\frac{1}{\lambda_{tr}\Gamma_0}\lb(\frac{\Phi}{1-\Phi}\rb).
\end{align}
Furthermore, $I_2$ can be approximated as
\begin{align} \label{I2_STS_asym}
    I_2\approx \int_{0}^{\infty}\frac{1}{(x+a)(x+b)^2}dx
\end{align}
where $a=\frac{n(\rho-1)\lambda_{sd}+\lambda_{sr}\lambda_{td}\xi}{n\rho\lambda_{sd}}$ and $b=\frac{\lambda_{te}\lambda_{sr}\xi}{\lambda_{se}}$. On substituting (\ref{xi_STS_asym}) and the solution of  (\ref{I2_STS_asym}) through the partial fraction method into (\ref{SOP_STS_asym}), the final asymptotic SOP can be derived as
\begin{align}
\label{eq_sts_asymp}
\mathcal{P}_{out}&\approx1-\sum_{n=1}^{N}\binom{N}{n}(-1)^{n+1}\frac{\lambda_{te}\lambda_{td}\lambda_{sr}^2\xi^2}{n\rho\lambda_{sd}\lambda_{se}}s^n \nn \\
&\times\lb(-\frac{\ln a}{(a-b)^2}+\frac{\ln b}{(a-b)^2}+\frac{1}{b(a-b)}\rb)
\end{align}

\subsection{Optimal Scheme for Transmitter Selection (OTS)}
By substituting $\Gamma_T \rightarrow \infty$, the expression within brackets in (\ref{sop_OTS}) can be approximated as
\begin{align}
&1-\frac{s\lambda_{se}\left(\Gamma_Ty+1\right)}{\rho\lambda_{sd}\left(\Gamma_Tx+1\right)+\lambda_{se}\left(\Gamma_Ty+1\right)}\xd
&\times\exp\left(\frac{-\lambda_{sd}\left(\rho-1\right)\left(\Gamma_T x+1\right)}{\Gamma_S}\right)\nn\\
&=1-\frac{s\lambda_{se}y}{\rho\lambda_{sd}x+\lambda_{se}y}\exp\lb({\frac{-\lambda_{sd}(\rho-1) x}{\xi\lambda_{sr}}}\rb).
\end{align}
The above equation leads to a closed-form solution for the asymptotic SOP after using the binomial expansion in (\ref{sop_OTS}). Finally, the asymptotic SOP can be expressed as in (\ref{eq_os_asymp}). The integral solution in (\ref{eq_os_asymp}) is obtained following equation (2.5.3.2) of \cite{prudnikov}. 
\begin{table*}
\begin{align}
\label{eq_os_asymp}
\mathcal{P}_{out}&\approx
\int_0^\infty\int_0^\infty \sum_{n=0}^{N}\binom{N}{n}\lb[(-1)^{n}\lb(\frac{s\lambda_{se}y}{\rho\lambda_{sd}x+\lambda_{se}y}\rb)^n\exp\left(\frac{-\lambda_{sd}(\rho-1)nx}{\xi\lambda_{sr}}\right)\rb]\lambda_{td}e^{-\lambda_{td}x}\lambda_{te}e^{-\lambda_{te}y}dxdy \nn\\
&=1-\frac{s\lambda_{se}\lambda_{td}\lambda_{te}N}{\rho\lambda_{sd}}\lb(\frac{\ln\lb(\frac{\lambda_{te}}{ab}\rb)}{\lb(\lambda_{te}-ab\rb)^2}-\frac{1}{\lb(\lambda_{te}-ab\rb)\lambda_{te}}\rb)-\sum_{n=2}^{N}\binom{N}{n}(-1)^{n+1}s^{n}b^n\frac{\lambda_{td}\lambda_{te}}{(n-1)!}\nn \\
&\times\left[\sum_{k=1}^{n-1}\frac{(k-1)!(n-k)!\lb(-a\rb)^{n-k-1}}{b^k\lambda_{te}^{n-k+1}}+\frac{\lb(-a\rb)^{n-1}\Gamma(n+1)}{(n+1)\lambda_{te}^{n+1}} {}_{2}F_{1}\lb(n+1,1;n+2;\frac{\lambda_{te}-ab}{\lambda_{te}}\rb)\right],
\end{align}
where $a=\frac{\lambda_{sd}\lb(\rho-1\rb)n+\lambda_{td}\lambda_{sr}\xi}{\lambda_{sr}\xi}$, $b=\frac{\lambda_{se}}{\rho\lambda_{sd}}$,  $\xi$ is the same as in (\ref{xi_STS_asym}), and ${}_{2}F_{1}(\cdot)$ is the hypergeometric function \cite{prudnikov}.\\  
\hrule
\end{table*}

\textit{Remark:} A closer look at the asymptotic expression for the SOP of the STS and OTS scheme in (\ref{eq_sts_asymp}) and (\ref{eq_os_asymp}), respectively, reveals that these are independent of $P_T$. This means the SOP saturates to a constant value as $\Gamma_T$ increases towards infinity.

\section{Numerical Results}\label{section 5}
In this section we provide numerical results, including both analytical results and simulations. We assume the system parameters are: $\beta$ = 0.5, $R_{th}=0.5$ bits/s/Hz, and  $\lbrace1/\lambda_{tr}, 1/\lambda_{td}, 1/\lambda_{sd}, 1/\lambda_{sr}, 1/\lambda_{te},1/\lambda_{se}\rbrace=\lbrace 3, -6, 3, -3, 6,-3\rbrace$ dB respectively. It is also assumed that the same noise power $N_0$  affects all nodes. We use red colour to represent cases where the TS is carried out with the available knowledge of active backhaul links, while the black colour represents TS when such knowledge in unavailable as in \cite{Kundu_TVT19}. Asymptotes are drawn for the cases of available backhaul knowledge only. It can be seen from the figures that in all cases, the analytical and simulated results match perfectly, thus validating our analysis.

Fig. \ref{S} shows the SOP versus $\Gamma_T$ for two distinct values of the backhaul success probability, $s = 0.5$ and $s = 0.99$, with $\Phi =0.1$, $N = 6$ and the knowledge of backhaul activity being used for selection. In general, it is observed that the performance of OTS is the best in all conditions, as expected. Furthermore, it is observed that the SOP performance improves with an increase in the backhaul reliability  $s$. The key observation is that the proposed TS schemes with the available knowledge of active transmitters significantly outperform the SOP performance of the TS scheme of \cite{Kundu_TVT19}. In particular, the proposed STS scheme with available knowledge of active transmitters can outperform the OTS scheme without the knowledge of backhaul activity. It can also be observed that the performance gain with available active transmitter knowledge is higher when the backhaul reliability improves (i.e., for higher $s$).
\begin{figure}
\centering
\includegraphics[width=3.1in,height=2.1in]{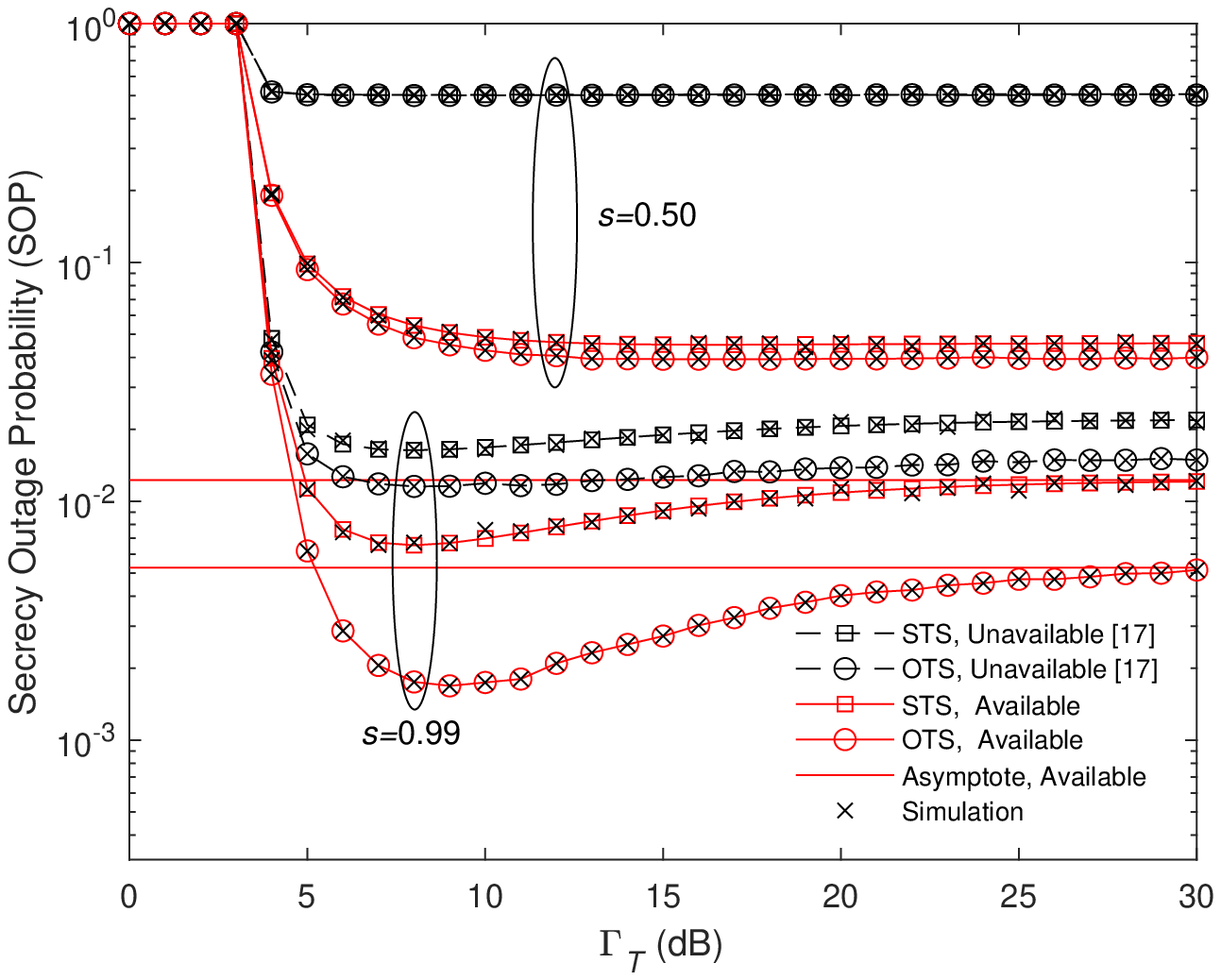}
 \vspace{-.25cm}
\caption{
SOP versus $\Gamma_T$(dB) for different values of $\textit{s}$.}
\label{S}
\vspace{-.57cm}
\end{figure}

 Fig. \ref{K} depicts the SOP versus $\Gamma_T$  for different numbers of secondary transmitters, $N = 2$ and $N = 6$, with parameters of the network set at $s = 0.99$ and $\Phi =0.1$. The SOP of the TS schemes improves as $N$ increases, irrespective of the available knowledge of active transmitters. This observation validates that  increasing the number of transmitters improves the available diversity of the system.
However, it is interesting to note that when the number of transmitters is higher ($N=6$), TS schemes with the knowledge of active backhaul links can provide higher performance gain. All other observations are similar to those of the previous figure.
 \begin{figure}
 \centering
 \includegraphics[width=3.1in,height=2.1in]{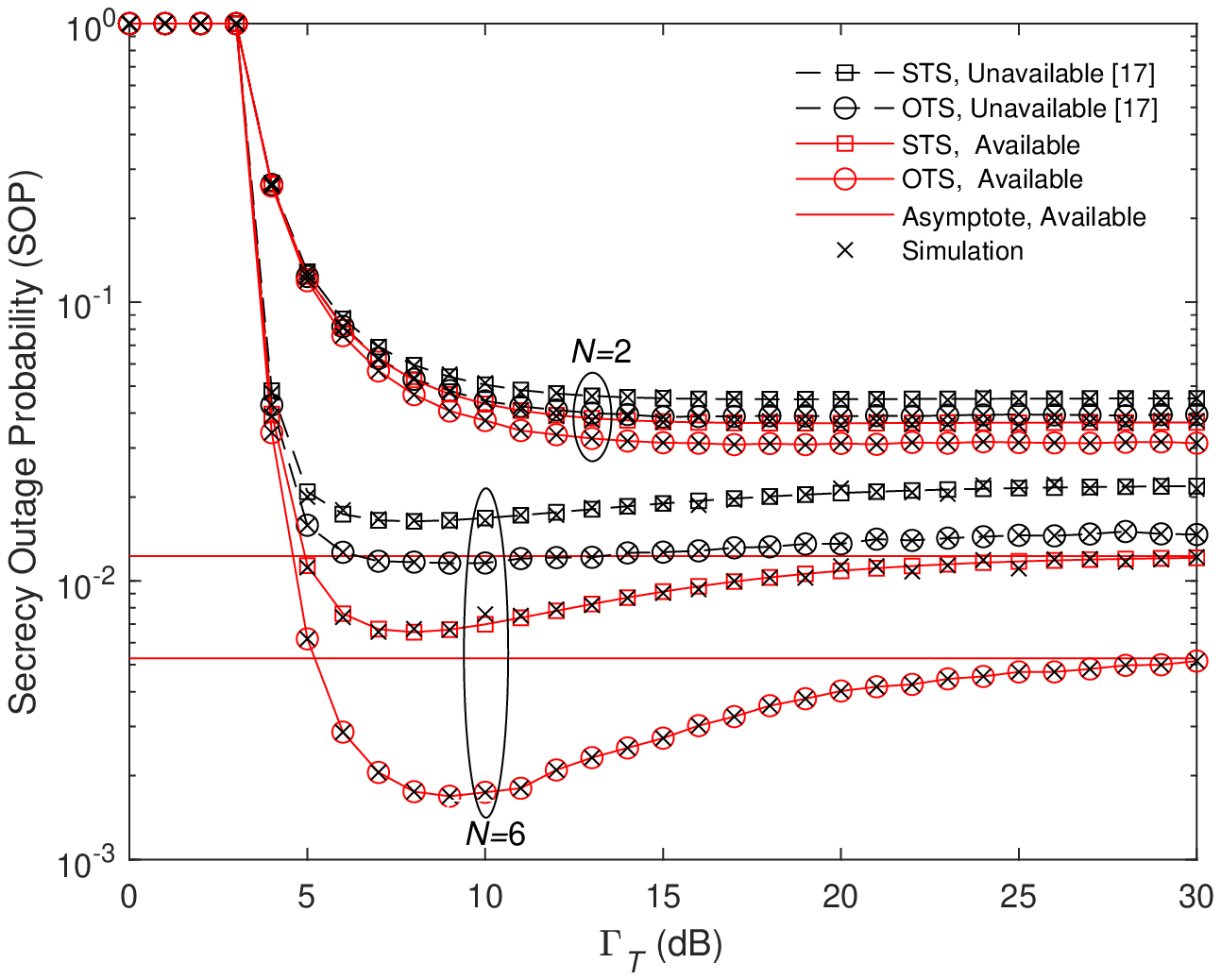} 
   \vspace{-.25cm}
 \caption{
 SOP versus $\Gamma_T$(dB) for different values of $\textit{N}$.}
 \label{K}
  \vspace{-.57cm}
 \end{figure}

In Fig. \ref{PHI}, the SOP versus $\Gamma_T$  is plotted for different values of the primary QoS constraint, $\Phi = 0.01$ and $\Phi = 0.1$, with network parameters $s = 0.99$ and $N =6$. It is observed that the SOP performance improves with increasing  $\Phi$ irrespective of whether backhaul activity knowledge is available or not. It can also be observed that as $\Phi$ increases from $\Phi=0.01$ to $\Phi=0.1$, TS schemes can provide better performance gain with the knowledge of active  backhaul.
Also, note that for lower values of $\Phi$ there is not much scope for SOP improvement with the knowledge of backhaul activity; this is due to the fact that the power constraint is more restrictive at lower $\Phi$.
\begin{figure}
 \centering
 \includegraphics[width=3.2in,height=2.1in]{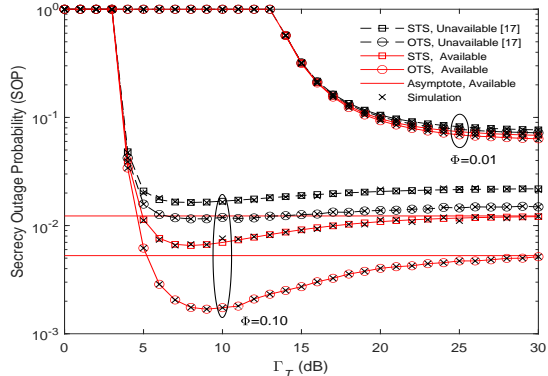} 
   \vspace{-.27cm}
 \caption{
 SOP versus $\Gamma_T$(dB) for different values of $\Phi$.}
 \label{PHI}
 \vspace{-.63cm}
 \end{figure}

A common observation from all the figures is that the SOP initially decreases below the asymptotic straight line and then subsequently increases towards this asymptotic value. This is because at higher values  of $\Gamma_T$, the power constraint $P_S$ at the secondary transmitter is not restrictive, however, at lower $\Gamma_T$ the power constraint becomes a significant issue for setting the secondary transmit power, which increases the SOP.
 \vspace{-.105cm}
  \section{Conclusion}\label{sec_Conclusion}
   \vspace{-.105cm}
 A sub-optimal and an optimal TS scheme have been proposed to improve the SOP performance of an underlay cognitive radio small-cell network operating with wireless backhaul. TS schemes are designed with the knowledge of backhaul activity along  with instantaneous channel state information at the transmitters. Closed-form analytical expressions (as well as asymptotic expressions) are  derived for the SOP of both selection schemes. Results show that a significant improvement in the SOP is achieved with the knowledge of active backhaul links, and the OTS scheme reaps the most additional benefit from this knowledge  compared to the case when knowledge of backhaul activity is unavailable.
  \vspace{-.105cm}
 \section*{Acknowledgment}
  \vspace{-.105cm}
 This work was supported in part by the Tata Consultancy Sevices (TCS) Foundation through its TCS Research Scholar Program, the Department of Science and Technology (DST) (Ref. No. DST/TMD/MI/OG/MI/2018/3(G)), the Science and Engineering Research Board (SERB), Government of India through its Early Career Research (ECR) Award (Ref. No. ECR/2016/001377) and  Science Foundation Ireland (SFI) under Grant Number 17/US/3445.
 \bibliographystyle{IEEEtran}
\bibliography{COG_backhaul}}
\end{document}